\documentclass[preprint,aps,superscriptaddress]{revtex4}
\usepackage[normalem]{ulem}
\usepackage{graphicx}
\usepackage{subfigure}
\usepackage{epsfig}
\usepackage{xcolor}
\usepackage{dcolumn}
\usepackage{bm}
\usepackage{ulem}
\usepackage{color}
\usepackage{amsmath}
\usepackage{amsfonts}
\usepackage{amssymb}
\usepackage{epstopdf}

\begin{document}

\title{Lorentz-violating type-II Dirac fermions in transition metal dichalcogenide PtTe$_2$}

\author{Mingzhe Yan}
\altaffiliation{These authors contribute equally to this work.}
\affiliation{State Key Laboratory of Low Dimensional Quantum Physics and Department of Physics, Tsinghua University, Beijing 100084, China}

\author{Huaqing Huang}
\altaffiliation{These authors contribute equally to this work.}
\affiliation{State Key Laboratory of Low Dimensional Quantum Physics and Department of Physics, Tsinghua University, Beijing 100084, China}

\author{Kenan Zhang}
\altaffiliation{These authors contribute equally to this work.}
\affiliation{State Key Laboratory of Low Dimensional Quantum Physics and Department of Physics, Tsinghua University, Beijing 100084, China}

\author{Eryin Wang}
\affiliation{State Key Laboratory of Low Dimensional Quantum Physics and Department of Physics, Tsinghua University, Beijing 100084, China}

\author{Wei Yao}
\affiliation{State Key Laboratory of Low Dimensional Quantum Physics and Department of Physics, Tsinghua University, Beijing 100084, China}

\author{Ke Deng}
\affiliation{State Key Laboratory of Low Dimensional Quantum Physics and Department of Physics, Tsinghua University, Beijing 100084, China}

\author{Guoliang Wan}
\affiliation{State Key Laboratory of Low Dimensional Quantum Physics and Department of Physics, Tsinghua University, Beijing 100084, China}

\author{Hongyun Zhang}
\affiliation{State Key Laboratory of Low Dimensional Quantum Physics and Department of Physics, Tsinghua University, Beijing 100084, China}

\author{Masashi Arita}
\affiliation{Hiroshima Synchrotron Radiation Center, Hiroshima University, Higashihiroshima, Hiroshima 739-0046, Japan}

\author{Haitao Yang}
\affiliation{State Key Laboratory of Low Dimensional Quantum Physics and Department of Physics, Tsinghua University, Beijing 100084, China}
\affiliation{Department of Physics and Tsinghua-Foxconn Nanotechnology Research Center, Tsinghua University, Beijing, 100084, China}

\author{Zhe Sun}
\affiliation{National Synchrotron Radiation Laboratory, University of Science and Technology of China, Hefei, Anhui 230029, China}

\author{Hong Yao}
\affiliation{Institute of Advanced Study, Tsinghua University, Beijing 100084, China}
\affiliation{Collaborative Innovation Center of Quantum Matter, Beijing, China}

\author{Yang Wu}
\altaffiliation{Correspondence should be sent to wuyang.thu@gmail.com, dwh@phys.tsinghua.edu.cn and syzhou@mail.tsinghua.edu.cn}
\affiliation{Department of Physics and Tsinghua-Foxconn Nanotechnology Research Center, Tsinghua University, Beijing, 100084, China}

\author{Shoushan Fan}
\affiliation{Department of Physics and Tsinghua-Foxconn Nanotechnology Research Center, Tsinghua University, Beijing, 100084, China}

\author{Wenhui Duan}
\altaffiliation{Correspondence should be sent to wuyang.thu@gmail.com, dwh@phys.tsinghua.edu.cn and syzhou@mail.tsinghua.edu.cn}
\affiliation{State Key Laboratory of Low Dimensional Quantum Physics and Department of Physics, Tsinghua University, Beijing 100084, China}

\author{Shuyun Zhou}
\altaffiliation{Correspondence should be sent to wuyang.thu@gmail.com, dwh@phys.tsinghua.edu.cn and syzhou@mail.tsinghua.edu.cn}
\affiliation{State Key Laboratory of Low Dimensional Quantum Physics and Department of Physics, Tsinghua University, Beijing 100084, China}
\affiliation{Collaborative Innovation Center of Quantum Matter, Beijing, China}

\date{\today}

\begin{abstract}

{\bf Topological semimetals have recently attracted extensive research interests \cite{ChenYLSci,HasanCdAs,HasanTaAs,DingHTaAs} as host materials to condensed matter physics counterparts of Dirac and Weyl fermions originally proposed in high energy physics. These fermions with linear dispersions near the Dirac or Weyl points obey Lorentz invariance, and the chiral anomaly leads to novel quantum phenomena such as negative magnetoresistance \cite{OngNaBi,OngCdAs,ChenTaAs}. The Lorentz invariance is, however, not necessarily respected in condensed matter physics, and thus Lorentz-violating type-II Dirac fermions with strongly tilted cones can be realized in topological semimetals \cite{BernevigNat,HasanLaAlGe}. Here, we report the first experimental evidence of type-II Dirac fermions in bulk stoichiometric PtTe$_2$ single crystal. Angle-resolved photoemission spectroscopy (ARPES) measurements and first-principles calculations reveal a pair of strongly tilted Dirac cones along the $\Gamma$-A direction under the symmetry protection, confirming PtTe$_2$ as a type-II Dirac semimetal. The realization of type-II Dirac fermions opens a new door for exotic physical properties distinguished from type-I Dirac fermions in condensed matter materials. }
\end{abstract}

\maketitle

In three dimensional topological Dirac semimetals, the low energy quasiparticle excitations are fermions described by the massless Dirac equation \cite{YoungSMPRL}, and such fermions are protected by certain crystalline symmetries in addition to time-reversal and inversion symmetries \cite{NagaosaNatcomm,WangzjNaBi,MorimotoPRB}. The generic Hamiltonian for a Dirac fermion is $H(\vec{k})=\sum\limits_{i=x,y,z\atop j=0,x,y,z}k_iA_{ij}\sigma_j$,where $\vec{k}$ is wave vector in momentum space, $\sigma_0$ is $2\times 2$ unit matrix, and $\sigma_j$ (\emph{j=x,y,z}) are the three Pauli matrices. The dispersion  $\epsilon_{\pm}(\vec{k})=\sum\limits_{i=x,y,z} k_iA_{i0}\pm \sqrt{\sum\limits_{j=x,y,z}(\sum\limits_{i=x,y,z}k_iA_{ij})^2}=T(\vec{k}) \pm U(\vec{k})$ are doubly degenerate as required by the presence of time-reversal and inversion symmetries \cite{BernevigNat}.  The degeneracy will be lifted when time-reversal or inversion symmetry is broken, and each Dirac fermion splits into a pair of Weyl fermions with opposite chiralities. The linear term $T(\vec{k})$ tilts the Dirac cone and the relative maganitude of $T(\vec{k})$ and $U(\vec{k})$ can be used to classify the topological nature of the Dirac or Weyl semimetals \cite{BernevigNat,HasanLaAlGe}. For type-I Dirac \cite{ChenYLSci,HasanCdAs} and Weyl semimetals \cite{HasanTaAs,DingHTaAs}, $T(\vec{k})$ is negligible compared to $U(\vec{k})$, and  massless Dirac cones with linear dispersions (see schematic drawing in  Fig.1(a)) are observed with isolated Dirac or Weyl points at the Fermi energy.  When $T(\vec{k})>U(\vec{k})$ along a certain direction, the Dirac cone is strongly tilted (Fig.~1(b)), leading to an intrinsic Lorentz violation. The fermions in this case emerge at the topologically protected points between electron and hole pockets, and there are finite density of states at the Fermi energy.  The different band topology also leads to distinguished magnetotransport properties. For example, while type-I Dirac and Weyl semimetals exhibit negative magnetoresistance along all directions \cite{OngNaBi,OngCdAs,ChenTaAs}, the magnetotransport in type-II semimetals is expected to be extremely anisotropic \cite{BernevigNat,CavaWTe,Zyuzin}.
Although type-II Weyl fermions have been reported recently \cite{ZhousyMT,KaminskiMT,HasanMT,ZhouxjMT,ChenYLMT,ShiMT,BaumbergerMT}, their spin-degenerate counterparts - type-II Dirac fermions still remain to be realized. Here by combining angle-resolved photoemission spectroscopy (ARPES) and first-principles calculations, we report the first discovery of Lorentz-violating type-II Dirac fermions in a transition metal dichalcogenide PtTe$_2$.

\begin{figure*}
\centering
\includegraphics[width=16.8 cm] {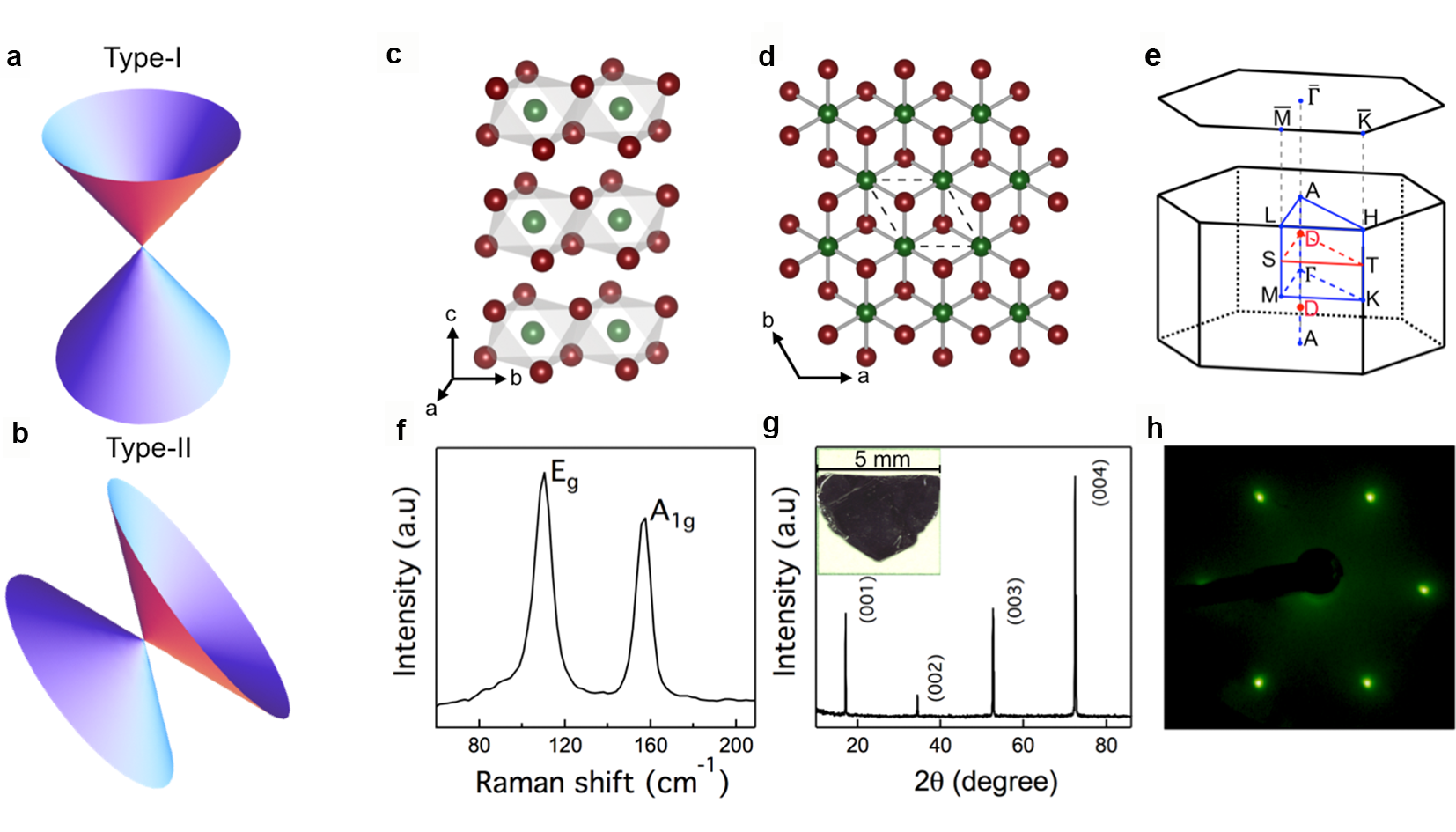}
\label{Figure 1}
\caption{{\bf Characterization of type-II Dirac semimetal PtTe$_2$.} \textbf{(a,b)}   Schematic drawing of type-I and type-II Dirac fermions. \textbf{(c-d)} Side and top views of the crystal structure of PtTe$_2$. Green balls are Pt atoms and red balls are Te atoms.  The unit cell is indicated by black dashed line.  \textbf{(e)}  Bulk and projected surface Brillouin zone along the (001) plane. Red dots (labeled as ``D'') mark the positions of the 3D Dirac points. \textbf{(f)} Raman spectrum measured at room temperature.  \textbf{(g)} XRD of PtTe$_2$ measured at room temperature. The inset shows the picture of one single crystal with a few mm size.  \textbf{(h)} LEED pattern taken at beam energy of 70 eV.}
\end{figure*}

PtTe$_2$ crystallizes in the CdI$_2$-type trigonal (1T) structure with \textit{P}$\bar{3}$\textit{m}1 space group (No. 162). The crystal structure is composed of edge-shared PtTe$_6$ octahedra with PtTe$_2$ layers tiling the ab plane (see Fig.~1(c,d)). The isostructural PtSe$_2$ monolayer film has been found to be a semiconductor with a gap of 1.2 eV \cite{PtSe2NL} and exhibits helical spin texture with spin-layer locking  induced by the local dipole field induced R-2 Rashba effect \cite{PtSe2spin}. Here we focus on the topological property of the semimetallic PtTe$_2$ bulk crystal, and we note that similar topological property is also expected in bulk PtSe$_2$. Figure 1(e) shows the hexagonal bulk Brillouin zone (BZ) and projected surface BZ onto the (001) surface. The Raman spectrum in Fig.~1(f) shows the E$_g$ and A$_{1g}$ vibrational modes at $\sim$ 110 cm$^{-1}$ and 157 cm$^{-1}$  respectively, which are typical for 1T structure \cite{Glamazda}. The sharp X-ray diffraction peaks (Fig.1(g)) and low-energy electron diffraction (LEED) pattern (Fig.~1(h)) confirm the high quality of the single crystals.

\begin{figure*}
\centering
\includegraphics[width=16.8 cm] {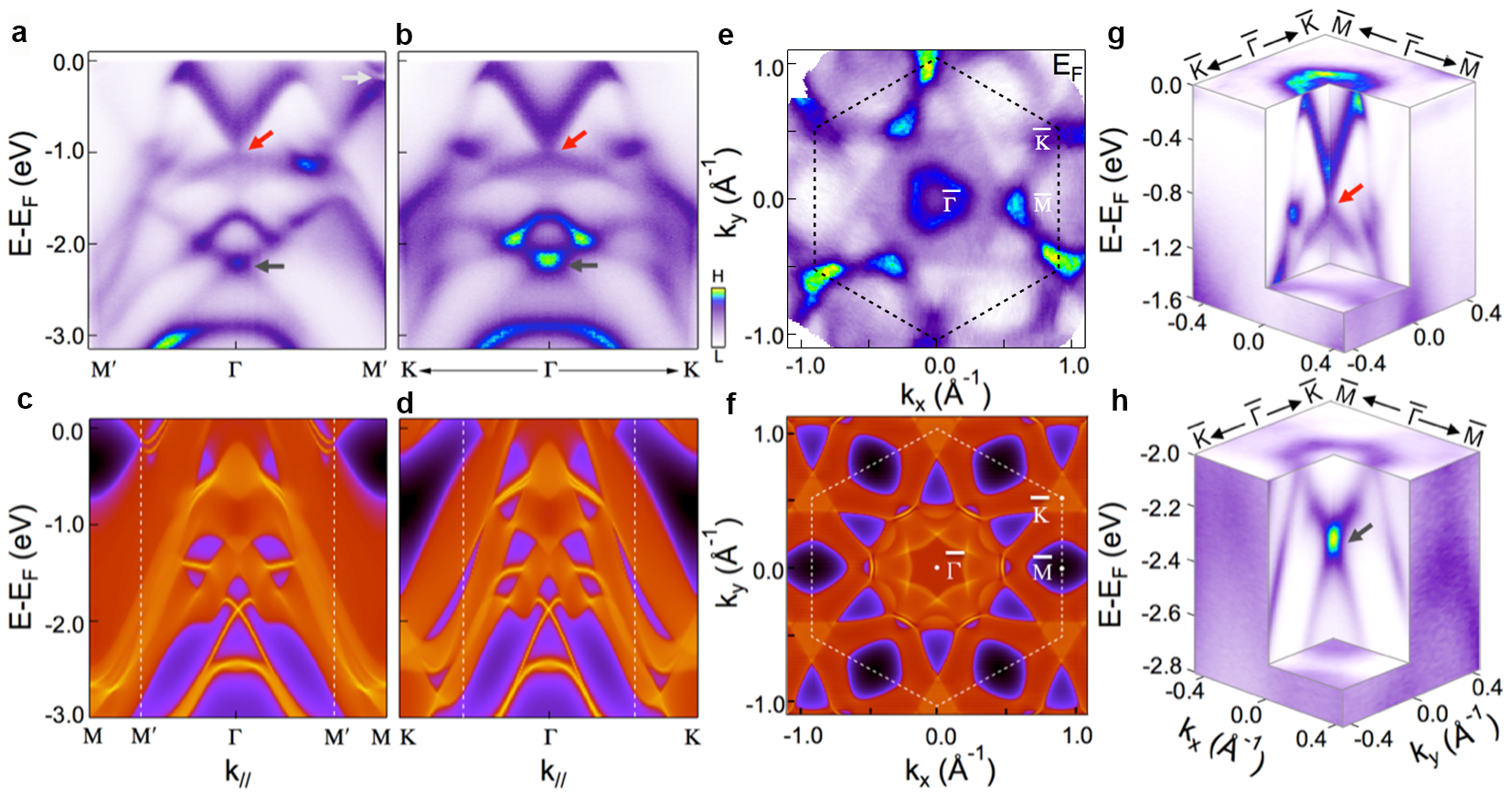}
\label{Figure 2}
\caption{{\bf The electronic structure of PtTe$_2$.}  \textbf{(a-b)} Band dispersions along the $\Gamma$-M (a) and $\Gamma$-K (b) direction at photon energy of 22 eV. \textbf{(c-d)} Calculated band dispersion along the $\Gamma$-M (c) and $\Gamma$-K (d) directions using Wannier function.  \textbf{(e)} Measured Fermi surface map at photon energy of 21.2 eV. Black dashed line indicates surface Brillouin zone.  \textbf{(f)} Calculated Fermi surface map. \textbf{(g-h)} Three-dimensional E-k$_x$-k$_y$ plots of the bulk Dirac cone around the $\Gamma$ point (g) and the surface state Dirac cone (h). }
\end{figure*}

The overview of the band structure of PtTe$_2$ near the fermi energy E$_F$ is shown in Fig.~2.  Figure 2(a,b) show ARPES data taken along two high symmetry directions $\Gamma$-M and $\Gamma$-K at photon energy of 22 eV. There are a few conical dispersions in the ARPES data. The most obvious one is centered at the $\Gamma$ point, which is formed by a V-shaped dispersion touching a flatter $\Lambda$-shaped dispersion (pointed by red arrow). The calculated projected spectral weight along the two high symmetric directions is shown in Fig.~2(c,d) for comparison. The cone-like dispersion discussed above corresponds to continuous states in the calculation, suggesting that this band is from the bulk states. This cone shows up as a pocket around the $\Gamma$ point in the measured and calculated Fermi surface maps (Fig.~2(e,f)). It is also clearly revealed in the three dimensional electronic structure shown in Fig.~2(g). The evolution of this cone with the out-of-plane momentum k$_z$ and its topological property are the main focus of this work.  The second conical dispersion is located between the $\Gamma$ and M points (labeled as M$^\prime$), and it is gapped at the Dirac point slightly below E$_F$ (pointed by white arrow). Calculated dispersions (Fig.~2(c)) show that this cone has bulk properties, and there are sharp surface states connecting the gapped Dirac cone here. The third conical dispersion is at much deeper energy between -2.0 eV and -2.6 eV (pointed by gray arrow in Fig.~2(a,b)). This cone corresponds to sharp surface states in the calculated dispersion (Fig.~2(c)). This cone connects the gapped bulk bands, similar to the Z$_2$ topological surface states observed in PdTe$_2$ \cite {ZhouXJCPL}.  The comparison between the measured and calculated band structure shows a good agreement with multiple conical dispersions arising from both the bulk bands and surface states.

\begin{figure*}
\centering
\includegraphics[width=16.8cm] {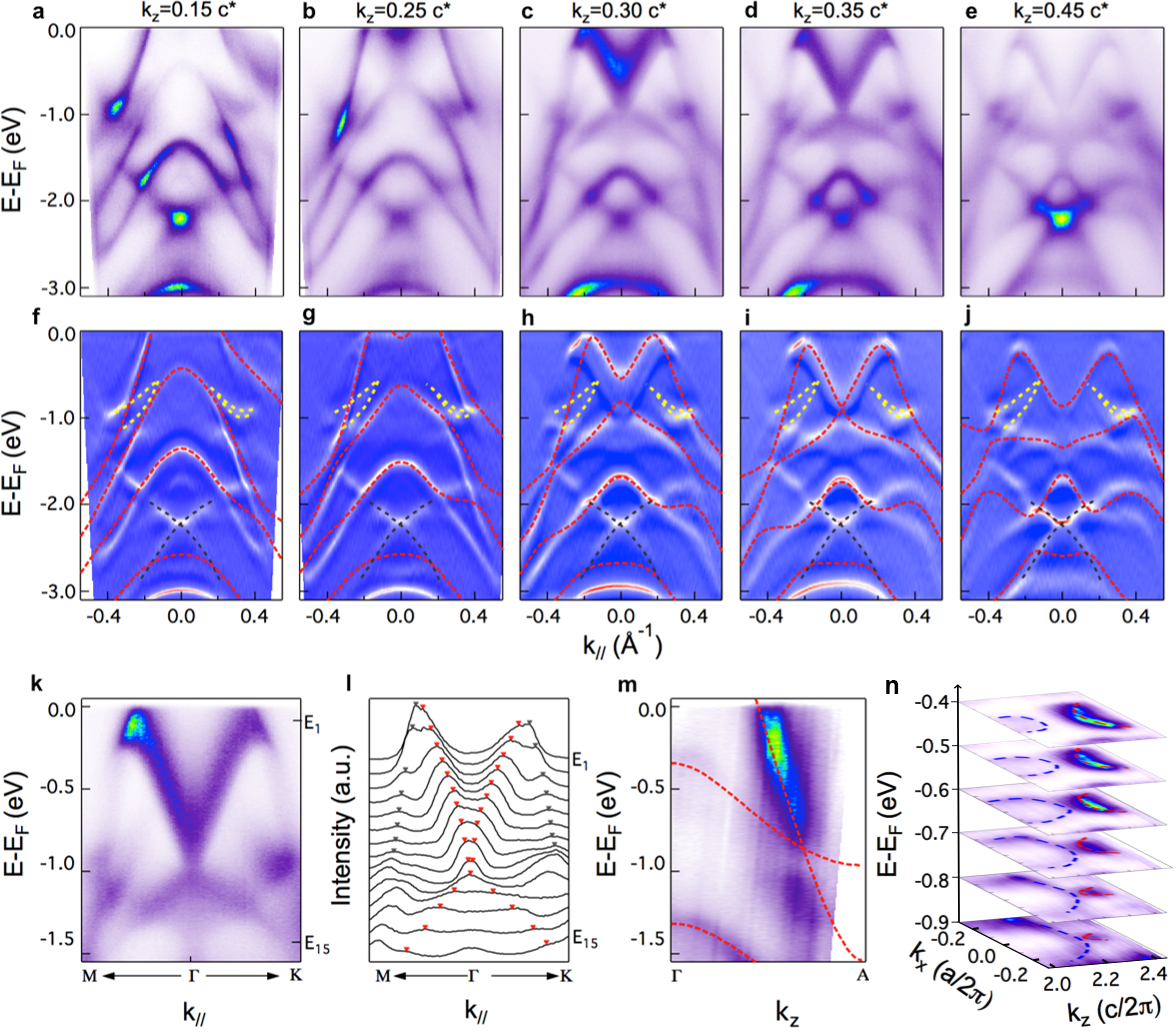}
\label{Fig3}
\caption{{\bf Evidence of type-II Dirac fermions in PtTe$_2$.}  \textbf{(a-e)} Dispersion along the M-$\Gamma$-K  direction measured at 17, 19, 21, 22 and 24 eV respectively. The corresponding k$_z$ values in the reduced BZ are labeled in each panel.  \textbf{(f-j)} Comparison of EDC-curvature with calculated dispersions from first-principles calculations (red dashed lines). Surface states are highlighted by yellow and gray dashed line respectively. \textbf{(k)} In-plane Dirac cone along the M-$\Gamma$-K direction measured at 22 eV. \textbf{(l)} MDCs for data shown in panel (k). Red and gray markers are guides for the peak positions of inner and outer bands. \textbf{(m)} Measured dispersion at k$_\parallel$ = 0. Red broken lines are calculated dispersions for comparison. There is a suppression of intensity around the band crossing, indicating a small gap. This could be explained by the small deviatation from the Dirac point, or strain induced symmetry breaking. \textbf{(n)}  Three dimensional map E - k$_x$ - k$_z$ at k$_y$=0, blue and red dashed line are guides for hole and electron pockets. The k$_z$ values are calculated in the extended Brillouin zone.}
\end{figure*}

To reveal the bulk versus surface properties of these Dirac cones, we show in Fig.~3 ARPES data measured along the $\Gamma$-K and $\Gamma$-M directions at different photon energies. The corresponding k$_z$ values are obtained using an inner potential of 13 eV, determined by comparing the experimental data with theoretical calculation.  Figure 3(a-e) shows the measured dispersions. The calculated bulk band dispersions at each k$_z$ value are overplotted on the curvature image in Fig.~3(f-j). A good agreement is obtained for the bulk Dirac cone at the $\Gamma$ point and its evolution with k$_z$.  The Dirac point discussed above is at k$_z$ $\approx$ 0.35 c$^*$, which is labeled as ``D" in Fig.~1(d). Away from this special point along the $\Gamma$-A direction, the valence and conduction bands begin to separate, and the separation becomes larger as k$_z$ moves further away from 0.35 c$^*$. The strong k$_z$ dependence confirms its three-dimensional nature. In addition to the bulk bands discussed above, there are surface states between -0.5 eV to -1 eV at the BZ center (highlighted by yellow dashed line in Fig.~3(f-j)) and at deeper energy (gray dashed line) which do not change with photon energy.

Figure 3(k) shows the zoom-in dispersion at the Dirac point. The conical dispersion can be clearly observed by following the peaks in the momentum distribution curves (MDCs) in Fig.~3(l). The type-II characteristics are revealed by plotting the dispersion as a function of k$_z$ (Fig.~3(m)) where a strongly tilted Dirac cone at the D point is revealed. The type-II characteristics is also reflected in the constant energy contours (Fig.~3(n)).  Three dimensional intensity map E-k$_x$-k$_z$ shows an electron pocket (red dashed lines) and a hole pocket (blue dashed lines) touching at the D point, which is another important signature of type-II Dirac cones. This anisotropic touching between the electron and hole pockets contributes finite density of states around the Dirac point and this is distinguished from the vanishing density of states at the Dirac point in type-I Dirac semimetal.

\begin{figure*}
\centering
\includegraphics[width=16.8 cm] {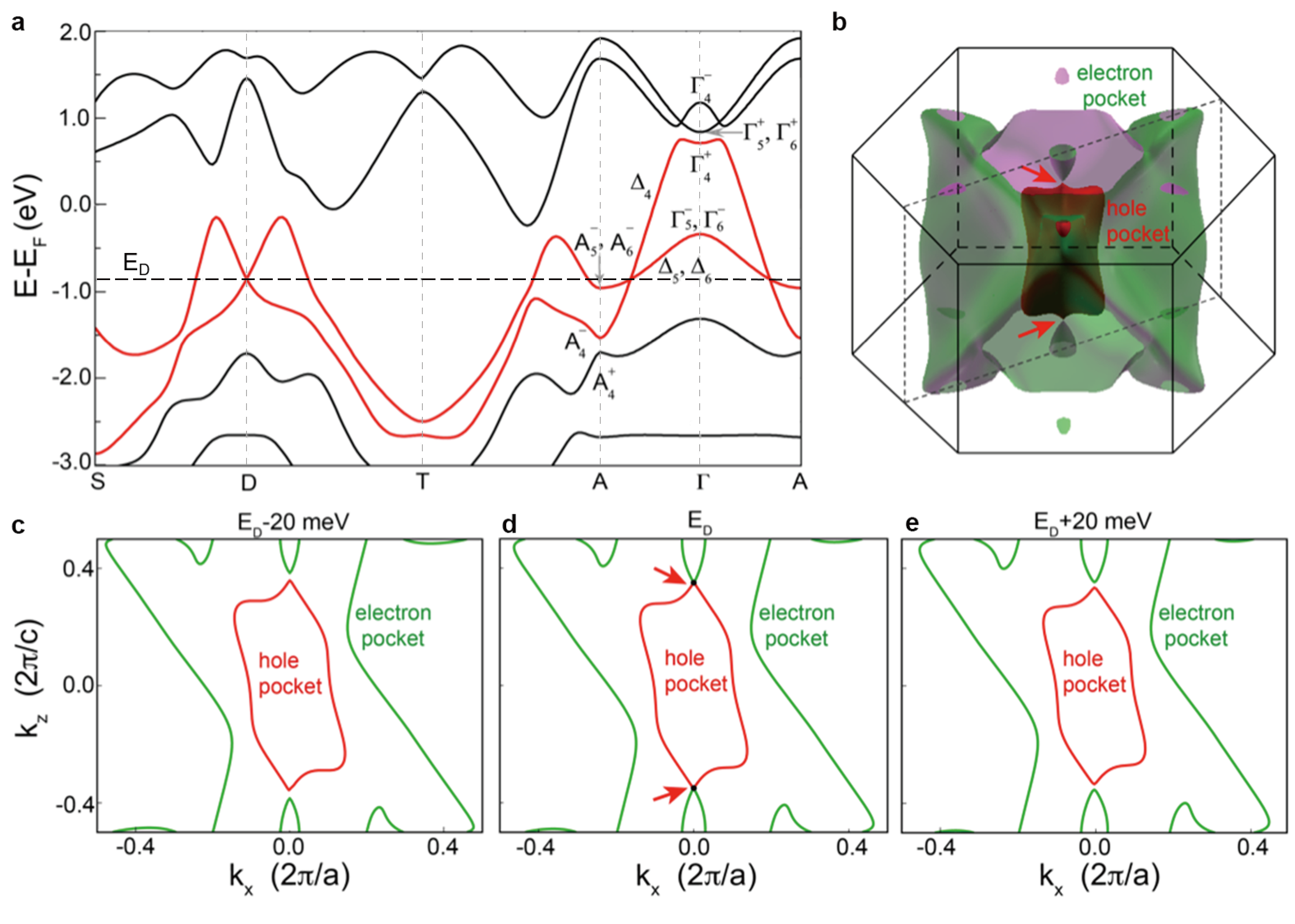}
\label{Fig4}
\caption{{\bf Theoretical calculation of type-II Dirac cone in PtTe$_2$.} \textbf{(a)} Calculated band dispersion along the in-plane direction S-D-T and out-of-plane direction A-D-$\Gamma$-D-A through the Dirac point. \textbf{(b)} Three dimensional plot of the electron and hole pockets at the Dirac point energy. The electron and hole pockets touch at the Dirac point D. \textbf{(c-e)} Contours of electron and hole pockets at -20, 0, +20 meV around the Dirac point energy in the k$_x$-k$_z$ plane. The electron and hole pockets touch at E$_D$ and clearly separated at E$_D$ $\pm$ 20 meV.}
\end{figure*}

In order to further reveal the topological nature of surface states and type-II Dirac cone in PtTe$_2$, we present the first-principles calculations of electronic structure and perform symmetry analysis. Figure 4(a) shows the calculated band structure along both the in-plane S-D-T and out-of-plane A-D-$\Gamma$ directions through the D point. Due to the band inversion between $\Gamma_4^+$ and $\Gamma_4^-$ at the $\Gamma$ point, there is a topologically nontrivial gap between them which gives rise to the surface states connecting the gapped cone structure at the M$^\prime$ point. In addition, there is another band inversion at the $A$ point at $\sim -2.5$ eV, which leads to the existence of the deep surface state as mentioned above. We have calculated the $Z_2$ invariants for bands below these two gaps, confirming the nontrivial topology of them. The bulk Dirac cone is formed by two valence bands with Te-$p$ orbitals (highlighted by red color). These two bands show linear dispersions in the vicinity of D along both the in-plane (S-D-T) and out-of-plane ($\Gamma$-$A$) directions, confirming that it is a three-dimensional Dirac cone.
This band crossing is unavoidable, because these two bands belong to different representations ($\Delta_4$ and $\Delta_{5+6}$), respectively, as determined by the $C_3$ rotational symmetry about the c axis \cite{NagaosaNatcomm}. The different irreducible representations prohibit hybridization between them, resulting in the symmetry-protected band crossing at $D=(0,0,0.346 c^*)$. As each band is doubly degenerate, the band crossing forms the four-fold degenerate Dirac point. We also calculated the energy contours by tuning the chemical potential to the Dirac point, as shown in Fig.~4(b). It is clear that there is a hole pocket in the BZ center (red color), while the much more complicated electron pockets (green color) are composed of a large outer pocket and a small inner one. The hole pocket and the small electron pocket touch each other at two Dirac points as shown in the isoenergy counter in the $k_x$-$k_z$ plane (Fig.~4(d)). By tuning the chemical potential above or below $E_D$ (Fig.~4(c,e)), we find that the hole and electron pockets disconnect, confirming that they only touch at the Dirac point.

In conclusion, by combining both ARPES measurements and first-principles calculations, we provide first direct evidence for the realization of type-II 3D Dirac fermions in single crystal PtTe$_2$.  Such type-II Dirac fermions violate Lorentz invariance and do not have counterpart in high energy physics. The realization of type-II Dirac semimetal provides a new platform for investigating various intriguing properties different from their type-I analogues, e.g. anisotropic magneto-transport properties.

{\bf Methods}

{\bf Sample growth} High quality PtTe$_2$ single crystal was obtained by self-flux methods. High purity Pt granules (99.9\%, Alfa Aesar) and Te lump (99.9999\%, Alfa Aesar) at a molar ratio of 2:98, were loaded in a silica tube, which is flame-sealed in a vacuum of $\sim$ 1 Pa. The tube was heated at 700$^{\circ}$C for 48 hours to homogenize the starting materials. The reaction was then slowly cooled to 480$^{\circ}$C at 5$^{\circ}$C/h to crystallize PtTe$_2$ in Te flux. The excess Te was centrifuged isothermally after 2 days.

{\bf ARPES measurement} ARPES measurements were taken at BL13U of Hefei National Synchrotron Radiation Laboratory, BL9A of Hiroshima Synchrotron Radiation Center under the proposal No.15-A-26 and our home laboratory. The crystals were cleaved \textit{in-situ} and measured at a temperature of T$\approx$20 K in vacuum with a base pressure better than 1$\times$10$^{-10}$ Torr.

{\bf Theoretical calculation} All first-principles calculations are carried out within the framework of density-functional theory using the Perdew-Burke-Ernzerhof-type \cite{PBE} generalized gradient approximation for the exchange-correlation potential, which is implemented in the Vienna \textit{ab initio} simulation package \cite{VASP}. A $8\times8\times6$ grid of \textbf{k} points and a default plane-wave energy cutoff are adopted for the self-consistent field calculations. Spin-orbit coupling is taken into account in our calculations. We calculate the surface spectral function and Fermi surface using the surface Green's function method \cite{lopez} based on maximally localized Wannier functions \cite{wannier90} from first-principles calculations of bulk materials.

{\bf Acknowledgements}
This work is supported by the National Natural Science Foundation of China (Grant No. 11274191, 11334006 and 11427903) and Ministry of Science and Technology of China (Grant No. 2015CB921001 and 2016YFA0301004).

{\bf Author Contributions}
S.Z., Y.W. and W.D. designed the project. M.Y., K.Z., E.Y., W.Y., K.D., G.W., H.Z. and S.Z. performed the ARPES measurements and analyzed the ARPES data. H.H, W.D performed theoretical calculation, K.Z., Y.W. prepared the single crystals. H.Y. discussed the data. M.Y. and S.Z. wrote the manuscript, and all authors commented on the manuscript.

{\bf Competing financial interests}
The authors declare no competing financial interests.

\end{document}